\documentclass[twocolumn]{aastex631}

\usepackage{hyperref}
\usepackage{todonotes}
\usepackage{chemformula}
\usepackage{makecell}
\newcommand\um{\textmu m }
\newcommand\degrees{$^\circ$ }
\usepackage{xcolor}

\received{Jan 10, 2024}
\revised{Mar 14, 2024}
\accepted{Mar 28, 2024}

\submitjournal{PSJ}
\graphicspath{{./}{figures/}}

\begin{document}

\title{Pwyll and Manann\'{a}n Craters as a Laboratory for Constraining Irradiation Timescales on Europa}

\correspondingauthor{M. Ryleigh Davis}
\email{rdavis@caltech.edu}

\author[0000-0002-7451-4704]{M. Ryleigh Davis}
\affiliation{Division of Geological and Planetary Sciences, California Institute of Technology, Pasadena, CA 91125, USA}

\author[0000-0002-8255-0545]{Michael E. Brown}
\affiliation{Division of Geological and Planetary Sciences, California Institute of Technology, Pasadena, CA 91125, USA}

\begin{abstract}
We examine high spatial resolution Galileo/NIMS observations of the young ($\sim$ 1 My - $\sim$ 20 My) impact features, Pwyll and Manann\'{a}n craters, on Europa's trailing hemisphere in an effort to constrain irradiation timescales. We characterize their composition using a linear spectral modeling analysis and find that both craters and their ejecta are depleted in hydrated sulfuric acid relative to nearby older terrain. This suggests that the radiolytic sulfur cycle has not yet had enough time to build up an equilibrium concentration of \ch{H2SO4}, and places a strong lower limit of the age of the craters on the equilibrium timescale of the radiolytic sulfur cycle on Europa’s trailing hemisphere. Additionally, we find that the dark and red material seen in the craters and proximal ejecta of Pwyll and Manann\'{a}n show the spectroscopic signature of hydrated, presumably endogenic salts. This suggests that the irradiation-induced darkening and reddening of endogenic salts thought to occur on Europa's trailing hemisphere has already happened at Pwyll and Manann\'{a}n, thereby placing an upper limit on the timescale by which salts are irradiation reddened.
\end{abstract}


\section{Introduction}\label{sec:intro}

\begin{figure*}[ht!]
\includegraphics[width=\textwidth]{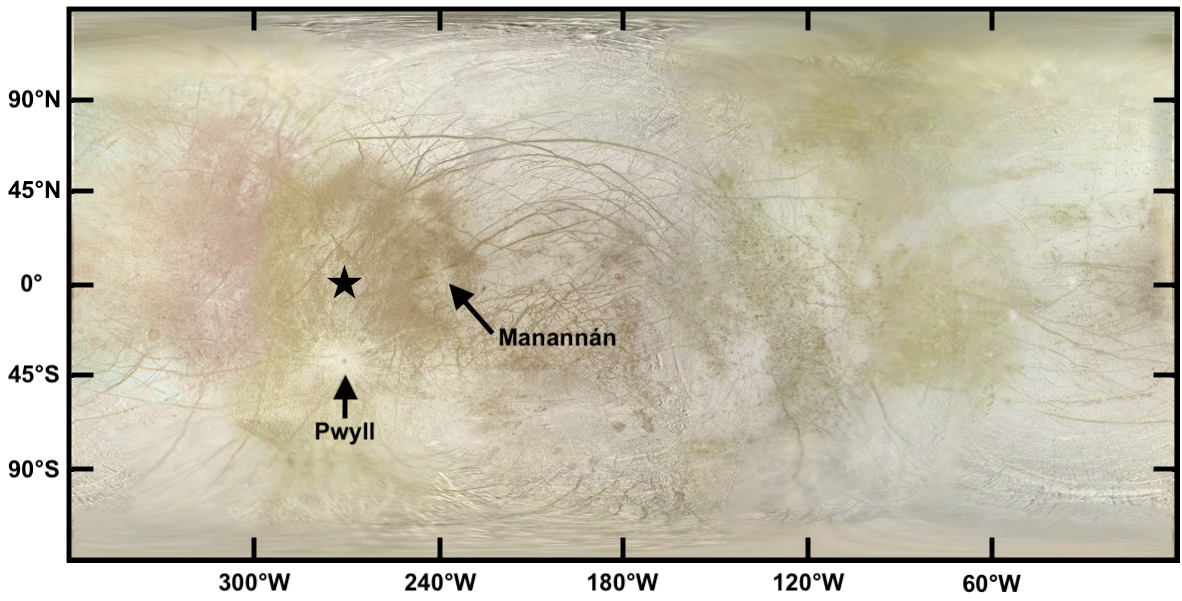}
\caption{Global mosaic of Europa showing the locations of Pwyll and Manann\'{a}n craters. Pwyll crater, one of Europa's youngest impact features, is $\sim$27 km in diameter, surrounded by a dark ring of proximal ejecta, and has an extensive bright ray ejecta which extends as far as 1000 km and stands in stark contrast to the darker reddish surface of Europa's trailing hemisphere. Manann\'{a}n crater is $\sim$23 km in diameter, also has a dark, red ring of proximal ejecta, and is surrounded by a much less extensive bright ray system which extends up to $\sim$120 km and covers nearby young terrain including chaos terrains in Dyfed Regio and the triple band Belus linea. The location of the trailing hemisphere apex (0$^\circ$ N, 270$^\circ$ W) is indicated with a star. This mosaic map, created by Steve Albers, blends an existing color map from NASA/JPL/Björn Jónsson created using color images from Voyager and Galileo, the high-resolution black and white mosaic from USGS, and additional higher resolution imagery from Galileo and Juno.
\label{fig:mosaic}}
\end{figure*}

Europa's trailing hemisphere surface composition reflects a complex interplay between endogenic material present within the recently geologically active chaos terrains and linea \citep[e.g.][]{mccord1999_HydratedSaltMinerals, dalton2005_SpectralComparisonHeavily, mccord2010_HydratedMineralsEuropa, dalton2012_LowTemperatureOptical, trumbo2020_EndogenicExogenicContributions}, exogenic material deposited by the Jovian magnetosphere \citep[e.g.][]{pospieszalska1989_MagnetosphericIonBombardment, cooper2001_EnergeticIonElectron, paranicas2001_ElectronBombardmentEuropa, paranicas2009_EuropaRadiationEnvironment}, and irradiation-induced alteration of both the endogenically and exogenically sourced material. \citep[e.g.][]{carlson1999_SulfuricAcidEuropa, brown2013_SALTSRADIATIONPRODUCTS, trumbo2020_EndogenicExogenicContributions}. 
Endogenic material within the chaos terrains and linea, thought to contain salts exhumed from Europa's subsurface \citep[e.g.][]{mccord1999_HydratedSaltMinerals, dalton2005_SpectralComparisonHeavily, mccord2010_HydratedMineralsEuropa, dalton2012_LowTemperatureOptical, trumbo2019_SodiumChlorideSurface, trumbo2022_NewUVSpectral}, appear to be irradiation darkened and reddened on Europa's trailing hemisphere, turning progressively redder toward the trailing hemisphere apex where the irradiation flux is strongest \citep{mcewen1986_ExogenicEndogenicAlbedo, nelson1986_EuropaCharacterizationInterpretation, johnson1988_AnalysisVoyagerImages, carlson2009_EuropaSurfaceComp}. A classic example of irradiation-induced alteration of exogenic material is the so-called radiolytic sulfur cycle, where Iogenic sulfur ions captured in Jupiter's magnetosphere are deposited on Europa's trailing hemisphere and subsequent irradiation produces hydrated sulfuric acid (\ch{H2SO4}) along with a variety of potential sulfur-bearing intermediary products in a ``bull's-eye'' pattern centered at the trailing hemisphere apex (0$^\circ$ N, 270$^\circ$ W) \citep{carlson1999_SulfuricAcidEuropa, carlson2002_SulfuricAcidProduction, carlson2005_DistributionHydrateEuropa}. Because of the intense irradiation environment on Europa's trailing hemisphere, these sulfur-bearing compounds are expected to be in radiation-induced equilibrium, with \ch{H2SO4} continuously produced and destroyed in some steady state concentration \citep{carlson1999_SulfuricAcidEuropa}. The timescale for reaching this irradiation-induced equilibrium on Europa's trailing hemisphere is not well known, however various estimates based on laboratory experiments have suggested timescales on the order of $\sim$10$^4$ yr \citep{carlson2002_SulfuricAcidProduction, strazzulla2007_HydrateSulfuricAcida, strazzulla2011_CosmicIonBombardment, ding2013_ImplantationMultiplyCharged}.

Young impact features may offer a unique opportunity for constraining irradiation timescales on Europa's trailing hemisphere. Pwyll crater and its ejecta, for example, show an enhancement in the UV-to-visible albedo ratio, which otherwise is anticorrelated with irradiation-induced discoloration on Europa's trailing hemisphere \citep{burnett2021_EuropaHemisphericColor}. \citet{burnett2021_EuropaHemisphericColor} suggest that the timescale for irradiation-induced discoloration on Europa must therefore be longer than the age of Pwyll crater. \citet{hendrix2011_EuropaDiskresolvedUltraviolet} also found that a 280 nm absorption feature attributed to \ch{SO2}, a likely product of the radiolytic sulfur cycle, had a low band strength near Pwyll, with the band strength generally increasing away from the crater and toward the older nearby chaos terrain. More recent work has revealed that Pwyll crater and its large ejecta blanket also lacks an unassigned absorption feature at 2.07 \um which appears to be formed via irradiation and is ubiquitous across the rest of Europa's trailing hemisphere \citep{davis2023_SpatialDistributionUnidentifieda}. 

\begin{figure}[ht!]
\plotone{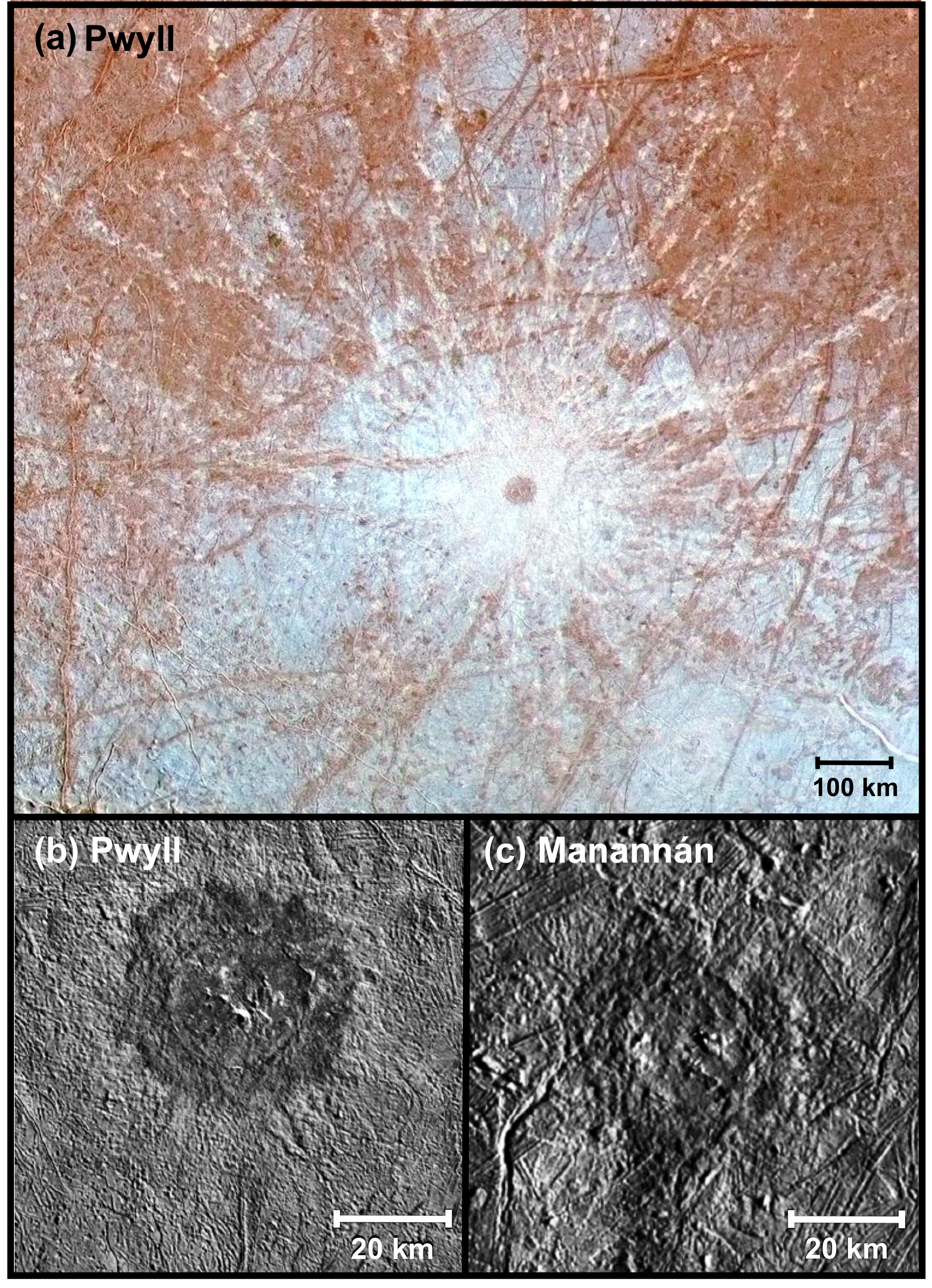}
\caption{(a) Enhanced color image constructed from Galileo/SSI images of Pwyll crater and the surrounding regions on Europa's trailing hemisphere. The dark, red crater and proximal ejecta are seen near the center of the image surrounded by an extensive bright ejecta blanket with rays extending over a thousand kilometers from the impact site. This bright ice-rich ejecta stands in stark contrast to the darker, reddish surface of Europa's trailing hemisphere. This image is from NASA/JPL/University of Arizona.
(b) Galileo/SSI image of Pwyll crater, taken with the clear filter. This image is from NASA/JPL/DLR.  (c) Galileo/SSI image of Manann\'{a}n crater taken with the clear filter. This image is from NASA/JPL/DLR.
In all three panels, north is towards the top of the image. For panels (b) and (c), the sun illuminates the surface from the right.
\label{fig:craters}}
\end{figure}

Pwyll and Manann\'{a}n craters are two of the youngest impact features on Europa. As can be seen in Figures \ref{fig:mosaic} and \ref{fig:craters} (a), Pwyll is a visually striking $\sim$27 km diameter impact feature on Europa's trailing hemisphere (25$^\circ$ S, 271$^\circ$ W) with a dark, reddish crater center surrounded by extensive bright white ejecta with rays that extend up to 1000 km from the impact site \citep{moore1998_LargeImpactFeatures}. It has a dark, red, pedestal-like proximal ejecta with an outward-facing steep slope or scarp which extends approximately half a crater diameter from the rim \citep{moore1998_LargeImpactFeatures} and is thought to contain endogenic material excavated from a depth of $\sim$1 km \citep{fanale2000_TyrePwyllGalileo} (see Figure \ref{fig:craters} (a), (b)). As can be seen in the global map of Europa (Figure \ref{fig:mosaic}), Pwyll is the only large crater which still has such an extensive ray system \citep{bierhaus2009crater}, which overlaps nearby chaos terrains in Dyfed Regio and Annwn Regio \citep{doggett2009geologic}, suggesting it is likely to be very young. Indeed, estimates for its age range from $\lesssim1 - 18$ Myr \citep{bierhaus2001_PwyllSecondariesOther, bierhaus2009crater}.

Manann\'{a}n crater (3$^\circ$ N, 240$^\circ$ W), shown in Figures 
\ref{fig:mosaic} and \ref{fig:craters} (c), has a crater diameter of $\sim$23 km \citep{moore1998_LargeImpactFeatures}. It has a partially preserved dark, red ring of proximal ejecta which may contain material excavated from Europa's subsurface that correlates with a pedestal-like slope break $\sim$7 km beyond the crater rim \citep{moore1998_LargeImpactFeatures}. Lower-lying regions on the crater floor are bluer than the surrounding material and proximal ejecta, suggesting an origin as impact melt \citep{moore2001_ImpactFeaturesEuropa, schenk2009impact}. The rim is generally weakly expressed and even absent in some areas \citep{schenk2009impact}, with evidence for lobate flows of material over the rim \citep{moore1998_LargeImpactFeatures, moore2001_ImpactFeaturesEuropa}.  Bright ray ejecta is seen extending up to $\sim$120 km from the impact, covering nearby chaos terrain in Dyfed regio as well as part of the nearby triple band Belus linea, suggesting that Manann\'{a}n is younger than all of the nearby terrains \citep{moore1998_LargeImpactFeatures}. However, based on ray preservation, Manann\'{a}n is expected to be older than Pwyll because at similar solar illumination, Manann\'{a}n's bright rays are much less prominent, significantly more degraded, and they cover a much smaller region \citep{moore2001_ImpactFeaturesEuropa, schenk2009impact}.

In this paper, we investigate the composition of the Pwyll and Manann\'{a}n impact features using existing high-spatial-resolution Galileo/Near-Infrared Mapping Spectrometer (NIMS) observations. We perform a simple linear spectral modeling analysis in order to better understand the composition of these craters and attempt to constrain irradiation timescales on Europa's trailing hemisphere. 

\section{Data Analysis}

We used the highest-spatial-resolution data available from Galileo/NIMS and associated images from the Solid State Imager (SSI) in order to examine the composition of Pwyll and Manann\'{a}n craters and the surrounding regions, with the aim of gaining insight into the various irradiation timescales operating on Europa's trailing hemisphere. In particular, we used a linear spectral modeling approach to explore the spatial distribution of water-ice, hydrated sulfuric acid, and salty ``non-ice'' material in and around each crater and its ejecta.

\subsection{Galileo/SSI}

\citet{bland2021_ImprovingUsabilityGalileo} recently updated the available Galileo/SSI dataset of Europa to account for uncertainties in the spacecraft's position and pointing which resulted in poor alignment of overlapping images, with some images displaced by more than 100 km from their correct location. We used the individual processed and projected ``level 2'' Galileo/SSI images with updated spatial information from \citet{bland2021_ImprovingUsabilityGalileo},  which we acquired from the USGS-hosted SpatioTemporal Asset Catalog (STAC, \href{https://stacspec.org/}{https://stacspec.org/}). We selected three images for our analysis. The first is a cutout from a larger image taken on 19 December 1996. We selected a region near Pwyll crater which covers an area $\sim$700 km wide with a spatial resolution $\sim$1.3 km/pixel and shows the low-albedo crater and proximal ejecta surrounded by an extensive blanket of bright ray ejecta. The second image, taken on 16 December 1997, shows a close-up view of Pwyll, including the dark crater and proximal ejecta at a spatial resolution of $\sim$150 m/pixel. The last image, taken on 29 March 1998, shows Manann\'{a}n crater, Belus linea, and the surrounding region at a spatial resolution of $\sim$1.65 km/pixel. All of the selected SSI images were taken with the clear filter at a relatively high solar illumination angle, which allowed for the best visible contrast between the dark material in the craters and proximal ejecta and the higher-albedo material of the bright ray systems. These SSI images can be seen in Figure \ref{fig:maps} (a), (d), and (g), respectively.

\subsection{Galileo/NIMS}
We used both of the existing high-spatial-resolution Galileo/NIMS observation of Pwyll crater and its ejecta. The first is NIMS cube 12ENCPWYLL01A taken on 16 December 1997 which covers the crater, the dark proximal ejecta, and only a very small amount of the surrounding bright ray ejecta at a spatial resolution of $\sim$1.5 km/pixel. We also used NIMS cube E6ENSUCOMP01A taken on 20 February 1997 which covers a larger area to the west of Pwyll crater and includes the bright ray ejecta and some older background terrain at a spatial resolution of $\sim$4.5 km/pixel. Finally, we used NIMS cube C3ENLINEA01B taken on 6 November 1996 which covers Manann\'{a}n crater, its ejecta, and much of the surrounding terrain, including Belus linea, at a spatial resolution of $\sim$11 km/pixel. The NIMS cube 12ENCPWYLL01A has been previously analyzed by \citet{fanale2000_TyrePwyllGalileo}, while the other two NIMS observations have remained unpublished. The data were downloaded from the NASA Planetary Data Archive \citep{gallileoNIMS} as calibrated I/F mosaics. From the Galileo/NIMS spectral image cubes PDS data set \citep{gallileoNIMS}, we used the individual PDS products E6ENSUCOMP01A-MSY02.IOF, 12ENCPWYLL01A-MSY02.IOF, and C3ENLINEA-01B-MSY04.IOF.

The included spatial information was calculated based on projection onto Europa according to spacecraft position, target position, and scan platform orientation, which are known to have relatively large uncertainties. We found that there were small offsets between the spatial information in the NIMS cubes and the recently updated SSI images from \citet{bland2021_ImprovingUsabilityGalileo}. We therefore applied small shifts in latitude and longitude to each of the NIMS data cubes to more closely match the SSI images. We empirically determined the optimal offsets using the results of our linear spectral modeling efforts (see section \ref{sec:specmod} below). In particular, we mapped the total water-ice abundance of each pixel where the higher water-ice abundances should correspond to the higher-albedo regions such as the bright ray ejecta in the SSI images \citep[e.g.][]{fanale2000_TyrePwyllGalileo} and the lower water-ice abundances correspond with the darker, less icy regions such as the crater interior, proximal ejecta, and older background terrains. The shapes of the bright ray ejecta are taken from \citet{leonard2024map}. We applied an offset of $-$0.25\degrees in latitude and $+$0.75\degrees in longitude for cube 12ENCPWYLL01A, $+$0.2\degrees in latitude and $-$0.25\degrees in longitude for cube E6ENSUCOMP01A, and $-$0.75\degrees in latitude and $-$0.75\degrees in longitude for cube C3ENLINEA01B. All of these applied offsets are small when compared with the field of view for their respective observations.

\subsection{Spectral Modeling}\label{sec:specmod}

\begin{table*}[]
    \centering
    \begin{tabular}{|c|c|c|c|c|}
    \hline
    \textbf{Species} & \textbf{Formula} & \textbf{Grain Size} &\textbf{ Temperature} & \textbf{Source}\\
    \hline \hline
     water-ice  & Crystalline  & 5, 50, 200, 1000 \um & 120 K & \citet{mastrapa2008_OpticalConstantsAmorphous}\\
     (Refractive indices) & Amorphous  & 5, 50, 200, 1000 \um & $>$70 K & \\
     \hline
     Sulfuric acid & \ch{H_2SO_4}$\cdot$ 6.5\ch{H_2O} & 50 \um & 80 K & \citet{carlson1999_SulfuricAcidEuropa}\\
     & \ch{H_2SO_4}$\cdot$ 8\ch{H_2O} & 5, 50 \um & 80 K & \\
     \hline
     Magnesium chloride & \ch{MgCl_2}$\cdot$ 2\ch{H_2O} & & 80 K & \citet{hanley2014_ReflectanceSpectraHydrated}\\
                        & \ch{MgCl_2}$\cdot$ 4\ch{H_2O} & & 80 K & \citet{hanley2014_ReflectanceSpectraHydrated}\\
                        & \ch{MgCl_2}$\cdot$ 6\ch{H_2O} & & 80 K & \citet{hanley2014_ReflectanceSpectraHydrated}\\
    \hline
    Magnesium chlorate & \ch{Mg(ClO_3)_2}$\cdot$ 6\ch{H_2O} & & 80 K & \citet{hanley2014_ReflectanceSpectraHydrated} \\
    \hline
    Magnesium perchlorate & \ch{Mg(ClO_4)_2}$\cdot$ 6\ch{H_2O} & & 80 K & \citet{hanley2014_ReflectanceSpectraHydrated}\\
    \hline
    Magnesium sulfate & \ch{MgSO_4} (\textit{aq}) & 50-100 \um & 100 K & \citet{dalton2005_SpectralComparisonHeavily} \\
    Hexahydrite       & \ch{MgSO_4}$\cdot$ 6\ch{H_2O} & 50-100 \um & 100 K & \citet{dalton2005_SpectralComparisonHeavily} \\
    Epsomite          & \ch{MgSO_4}$\cdot$ 7\ch{H_2O} & 25-200 \um & 120 K & \citet{dalton2012_LowTemperatureOptical}\\
    \hline
    Mirabilite &  \ch{Na_2SO_4}$\cdot$ 10\ch{H_2O} & 50-100 \um & 100 K & \citet{dalton2005_SpectralComparisonHeavily} \\
    \hline
    Sodium chloride & \ch{NaCl} & & 80 K & \citet{hanley2014_ReflectanceSpectraHydrated}\\
    \hline
    Sodium perchlorate & \ch{NaClO_4} & & 80 K & \citet{hanley2014_ReflectanceSpectraHydrated}\\
                       & \ch{NaClO_4}$\cdot$ 2\ch{H_2O} & & 80 K & \citet{hanley2014_ReflectanceSpectraHydrated}\\
    \hline
    \end{tabular}
    \caption{Laboratory spectra used in our linear spectral modeling analysis, based on the spectral library of \citet{king2022_CompositionalMappingEuropa}. When available, the temperature and grain size of each laboratory spectrum is listed.}
    \label{tab:lib}
\end{table*}

We performed a simple linear spectral modeling analysis to explore the relative contributions of water-ice, hydrated sulfuric acid, and salts to the spectra in and around Pwyll and Manann\'{a}n craters. We based our spectral library on that of \citet{king2022_CompositionalMappingEuropa}, who modeled much of Europa's surface from a combination of global Galileo/NIMS and Very Large Telescope/SPHERE observations. This spectral library includes crystalline and amorphous water-ice, several hydration states of sulfuric acid, and a variety of cryogenic salts. The specific laboratory spectra included in the library and their relevant properties are listed in Table \ref{tab:lib}. Water-ice spectra, for both crystalline and amorphous water-ice, were calculated using laboratory optical constants \citep{mastrapa2008_OpticalConstantsAmorphous} and the Shkuratov model \citep{shkuratov1999_ModelSpectralAlbedo} for grain sizes of 5 \um, 50 \um, 200 \um, and 1 mm and a volume filling fraction of 0.7.

Each NIMS spectrum was analyzed by fitting a linear combination of laboratory spectra from the reference library, where the model spectrum, $M_{\lambda}$, is described by 
\begin{equation}
    M_{\lambda} = C \sum_{i} w_i \cdot S_{i, \lambda} \ \ \ \text{where} \ \sum_{i} w_i = 1
\end{equation}
where $w_i$ are the weights associated with each laboratory end-member spectrum, $S_i$. The additional scalar scaling factor, C, accounts for small variations in the reflectance between the calibrated NIMS observations and the measured laboratory spectra. These differences may be due to photometric effects caused by differing illumination of the laboratory samples and Europa's surface, as well as errors associated with the use of a linear mixing model. Typical values of C range from $\sim$0.4-0.7 for NIMS cube 12ENCPWYLL01A, $\sim$0.3-0.5 for NIMS cube E6ENSUCOMP01A, and $\sim$0.8-1.2 for NIMS cube C3ENLINEA01B. We used the non-negative least squares algorithm implemented in SciPy v1.10.1 to solve for the best-fit spectral model at each NIMS pixel. 

We map the relative contribution of sulfuric acid and salts to the best-fit spectral models for each NIMS cube. We calculate the abundance of sulfuric acid in each pixel by summing the individual coefficients of all three laboratory sulfuric acid spectra, which have different grain sizes and hydration states. Similarly, we report the relative abundance of the ``salty" material by summing the coefficients of all of the salts included in our spectral library. While our library includes a variety of hydrated salts which are potentially present on Europa's surface, the specific composition of Europa's dark endogenic material is not well known.  The near-infrared spectra of Europa's dark hydrate material, as well as many of the candidate salt species, are mainly characterized by asymmetric distortions to the water-ice bands and lack specific characteristic absorptions. This lack of distinguishing spectral features, particularly at the low spectral resolution of NIMS, results in degeneracies between various salt endmembers, which make positive identification of specific species on Europa's surface difficult \citep[e.g.][]{king2022_CompositionalMappingEuropa}. We therefore choose not to report the specific coefficients for each individual salt, but rather only consider the total contribution of the non-water-ice and non-sulfuric-acid ``salty" hydrated material to the spectral model, hereafter referred to as the salt abundance. 

The measurement uncertainty in the NIMS observations is not well known, and the influence of non-quantifiable radiation noise means that a simple error estimate based on the inherent scatter of each spectrum cannot robustly predict the true error. We therefore adopt a Monte Carlo bootstrap method to estimate the error in our model abundances, where we use the idea of a constant chi-squared boundary to predict the 1$\sigma$ confidence limit for each of our model weights, as described in \citet{press1992_NumericalRecipes} section 15.6. We first use a bootstrap sampling method to generate 1000 synthetic spectra for each NIMS spectra, with the same number of data points as the original spectrum, by drawing N (wavelength, I/F) pairs with replacement from the original spectrum. We then use the same spectral library and chi-squared minimization algorithm as before to estimate a new set of model abundances for each synthetic spectrum and calculate a set of corresponding chi-squared values, with respect to the original NIMS spectra for which we want to find error bars. Each of these new chi-squared values, calculated for a set of model coefficients which have been perturbed away from the best-fit abundances, are necessarily larger than the minimum chi-squared for the original NIMS spectrum. As described in \citet{press1992_NumericalRecipes}, there is some region within which the $\chi^2$ value increases by no more than a set $\Delta \chi^2$ which defines an M-dimensional confidence region. We find the approximate location of the constant $\Delta \chi^2$ boundary that corresponds to the 1$\sigma$ confidence region by considering the range of model abundances which have a corresponding chi-squared value in the lowest 68\% of the set of chi-squared values. We then use the range of model abundances present in the set to estimate 1$\sigma$ error bars for each of the M model abundances corresponding to the specific library spectra, as well as the error bars for our combined compositional categories of water-ice, sulfuric acid, and salts. These estimated 1$\sigma$ error bars are reported throughout the paper along with the best-fit abundance values.

\section{Results}

\begin{figure}[ht!]
\plotone{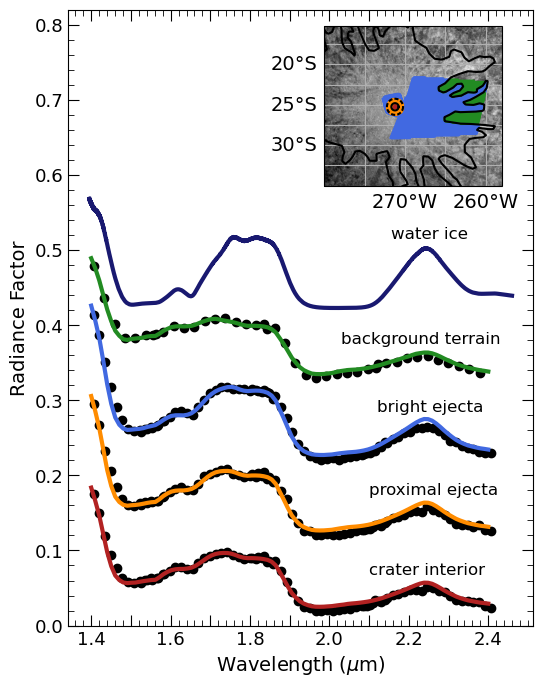}
\caption{Average spectra of Pwyll crater, its dark proximal ejecta, bright ray ejecta, and the nearby surrounding terrain along with their best-fit spectral models. The NIMS pixels included in the average spectra are shown in the inset map with the crater interior in red, the dark proximal ejecta in orange, the bright ray ejecta in blue, and the older background terrain in green. A laboratory derived spectrum of pure water-ice with 200 \textmu m grains from \citet{mastrapa2008_OpticalConstantsAmorphous} is also included for comparison. For clarity, the spectra are offset by +0.0, +0.1, +0.2, +0.3, and +0.4, respectively.
\label{fig:pwyll_specs}}
\end{figure}

\begin{figure}[ht!]
\plotone{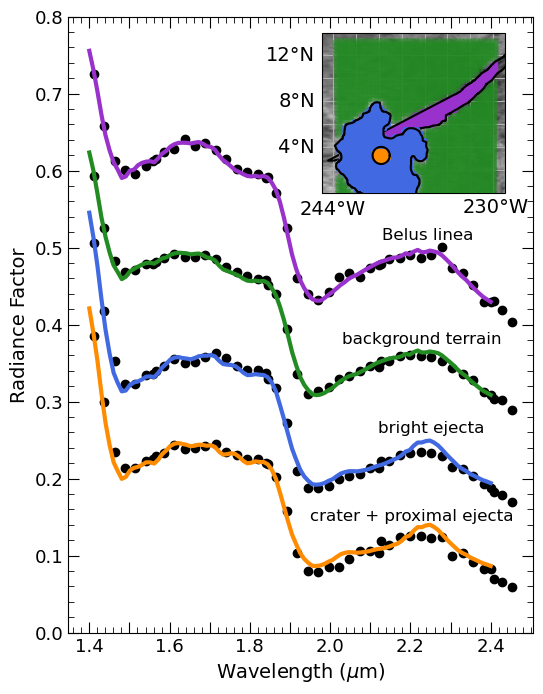}
\caption{Average spectra and best-fit spectral models of Manann\'{a}n crater and its dark proximal ejecta, bright ray ejecta, and the nearby surrounding terrain along with Belus linea. The NIMS pixels included in the average spectra are shown in the inset map with the crater interior and dark proximal ejecta in orange, the bright ray ejecta in blue, the older background terrain in green, and Belus linea in purple. For clarity, the spectra are offset by +0.0, +0.1, +0.2, and +0.3, respectively.
\label{fig:manannan_specs}}
\end{figure}

\begin{table}
\begin{flushleft}
\begin{tabular}{|l|ccc|}
    \hline
    & Sulfuric Acid & Salts & Water-Ice \\
    \hline
    \textbf{Pwyll (Figure \ref{fig:pwyll_specs}):} & & & \\
    \hspace{1.5mm} Crater interior & 24$^{+14}_{-14}$\% & 11$^{+15}_{-8}$\% & 65$^{+7}_{-26}$\% \\
    \hspace{1.5mm} Proximal ejecta & 22$^{+15}_{-14}$\% & 10$^{+12}_{-9}$\% & 68$^{+9}_{-21}$\% \\
    \hspace{1.5mm} Bright ray ejecta & 24$^{+13}_{-19}$\% & 5$^{+20}_{-5}$\% & 71$^{+12}_{-17}$\% \\
    \hspace{1.5mm} Background terrain & 44$^{+12}_{-29}$\% & 17$^{+26}_{-10}$\% & 39$^{+23}_{-25}$\% \\
    \textbf{Manann\'{a}n (Figure \ref{fig:manannan_specs}):} & & &\\
    \hspace{1.5mm} Crater+proximal ejecta & 41$^{+11}_{-22}$\% & 29$^{+32}_{-10}$\% & 30$^{+25}_{-17}$\%\\
    \hspace{1.5mm} Bright ray ejecta & 45$^{+11}_{-25}$\% & 26$^{+22}_{-24}$\% & 29$^{+24}_{-16}$\% \\
    \hspace{1.5mm} Background terrain & 55$^{+8}_{-21}$\% & 42$^{+7}_{-20}$\% & 3$^{+16}_{-3}$\% \\
    \hspace{1.5mm} Belus linea & 45$^{+19}_{-26}$\% & 47$^{+14}_{-19}$\% & 8$^{+27}_{-8}$\%\\
    \hline
\end{tabular}
    \caption{Best-fit spectral modeling results for the average spectra of specific compositional units of the Pwyll and Manann\'{a}n regions from Figures \ref{fig:pwyll_specs} and \ref{fig:manannan_specs}. The estimated 1$\sigma$ confidence region for the best-fit model abundances are also reported.}
    \label{tab:coeffs}
\end{flushleft}
\end{table}

\begin{figure}[ht!]
\plotone{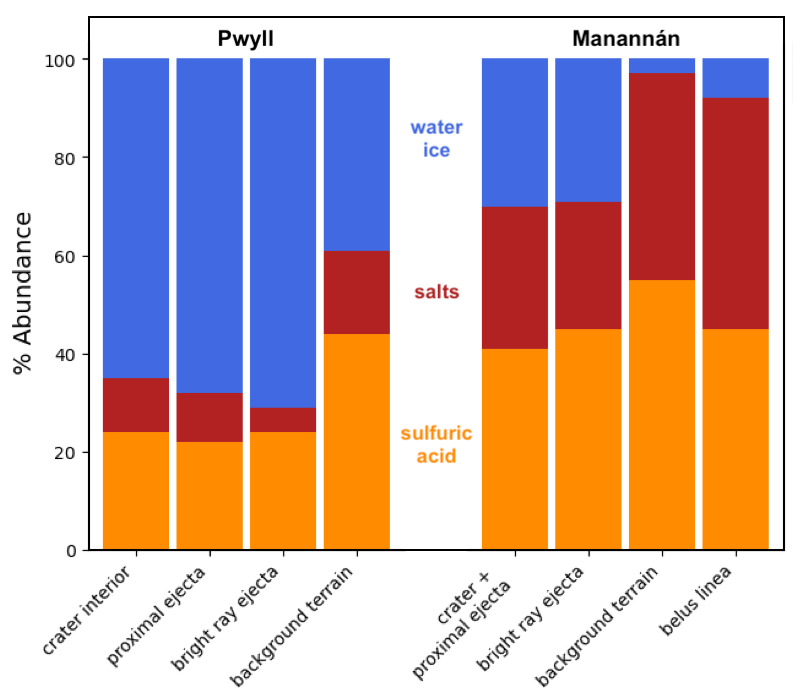}
\caption{Abundance values for the best-fit spectral models of the average spectra of specific geologic units in the Pwyll and Manann\'{a}n regions. These abundance values and estimated 1$\sigma$ error bars are reported in Table \ref{tab:coeffs} and correspond to the models shown in color in Figures \ref{fig:pwyll_specs} and \ref{fig:manannan_specs}. At both Pwyll and Manann\'{a}n, the average spectra of the crater, proximal ejecta, and bright ray ejecta are all depleted in sulfuric acid relative to the background terrain. The dark material in both of the craters and proximal ejecta contain a higher abundance of salts than the bright ray ejecta, although the background terrains appear to have the highest concentration of salty material. Manann\'{a}n crater and the surrounding region shows a higher concentration of both salts and sulfuric acid, and therefore less pure water-ice than Pwyll.
\label{fig:abundance_bars}}
\end{figure}

Figures \ref{fig:pwyll_specs} and \ref{fig:manannan_specs} show a few average NIMS spectra for specific compositional units along with their best-fit spectral models for Pwyll and Manann\'{a}n, respectively. The NIMS pixels included in each average spectra are shown in the inset maps. The best-fit model abundances for sulfuric acid, salts, and water-ice for each of these average spectra are reported in Table \ref{tab:coeffs} and shown graphically in Figure \ref{fig:abundance_bars}. The estimated 1$\sigma$ error bars are included in Table \ref{tab:coeffs}. We note that NIMS cube E6ENSUCOMP01A, which includes many of the bright ray ejecta pixels and all of the background terrain pixels near Pwyll, was taken with a lower spectral resolution ($\sim$0.026 \um) than the other cube which covers Pwyll crater and the proximal ejecta ($\sim$ 0.015 \um). To compute the average spectrum for Pwyll's bright ray ejecta, which includes pixels in both NIMS cubes, we interpolate the lower-resolution spectra to match the wavelengths of the higher-resolution observation.

In Figure \ref{fig:pwyll_specs}, the average spectra of the Pwyll crater interior and proximal ejecta show strong distortions to the water-ice absorption bands, particularly in the $\sim$2-2.2 \um range, which is indicative of hydrated, presumably endogenic salts while the bright ray ejecta most closely resembles a pure water-ice spectrum. The low-albedo material in the crater and proximal ejecta are apparently enriched in salts relative to the bright ray ejecta. Indeed, the best-fit spectral model for the average spectrum of the crater and proximal ejecta show an abundance of salts (11$^{+15}_{-8}$\% and 10$^{+12}_{-9}$\%) which is around twice that of the average spectrum of the bright ray ejecta (5$^{+20}_{-5}$\%). The spectral signature of hydrated sulfuric acid can be seen most prominently in the average spectrum of the older background terrain, which has a model abundance of 44$^{+12}_{-29}$\% sulfuric acid, as compared with 22$^{+15}_{-14}$ - 24$^{+13}_{-19}$\% sulfuric acid found in the average crater and ejecta spectra. This suggests that Pwyll crater and its ejecta are depleted in sulfuric acid relative to the older nearby terrain. 

The NIMS observations of Manann\'{a}n crater are at sufficiently low spatial resolution ($\sim$11 km/pixel), that it is difficult to distinguish between the crater interior and dark proximal ejecta. We therefore combine these units and consider the crater interior and proximal ejecta together. As seen in Figure \ref{fig:manannan_specs} and Figure \ref{fig:abundance_bars}, the average spectrum of the crater/proximal ejecta and bright ray ejecta of Manann\'{a}n appear to be somewhat depleted in sulfuric acid (41$^{+11}_{-22}$ and 45$^{+11}_{-25}$\%) relative to the nearby background terrain (55$^{+8}_{-21}$\%), although less significantly than at Pwyll crater. The average spectrum of the nearby Belus linea is also depleted in sulfuric acid when compared with the background terrain. Additionally, the average spectrum of the Manann\'{a}n crater and proximal ejecta, as well as the bright ray ejecta appear to contain a higher abundance of salts than is seen at Pwyll. At both Pwyll and Manann\'{a}n, the average background terrain spectra have a higher best-fit abundance of salt than the crater and ejecta material, although there is a significantly higher salt abundance seen in the background terrain near Manan\'{a}n (42$^{+7}_{-20}$\%) than Pwyll (17$^{+26}_{-10}$\%). Belus linea contains an even stronger salt signature (47$^{+14}_{-19}$\%) than the background terrain near Manan\'{a}n.

\begin{figure*}[ht!]
\includegraphics[width=\textwidth]{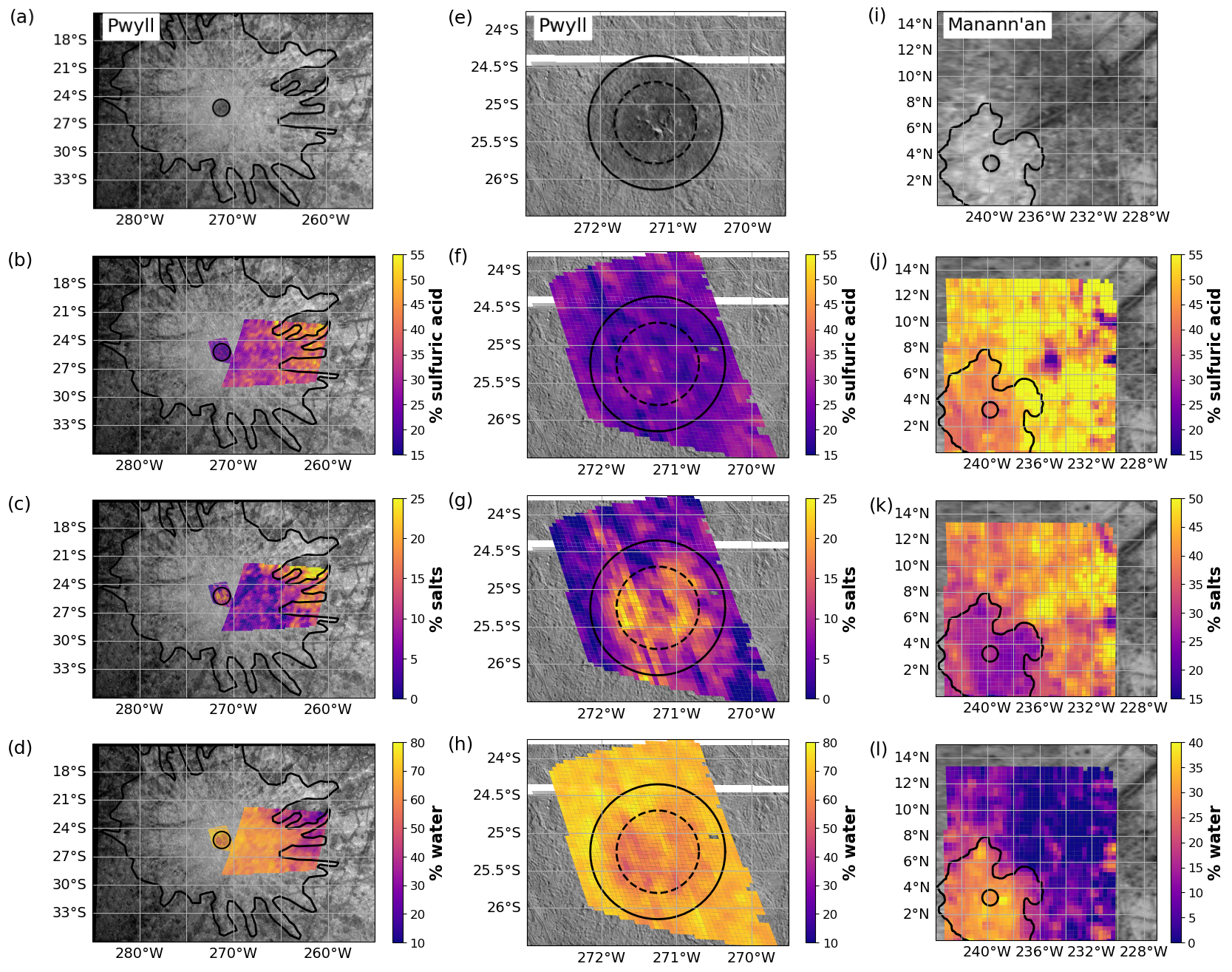}
\caption{Sulfuric acid, salt, and water-ice concentration maps for Pwyll and Manann\'{a}n craters. Panels (a) and (e) show the Galileo/SSI images of Pwyll crater, with the estimated sulfuric acid abundance from the corresponding NIMS cubes displayed in panels (b) and (f), the salt abundance in panels (c) and (g), and the water-ice abundance in panels (d) and (h). The crater rim is shown as the dashed circle, the extent of the dark proximal ejecta is indicated by the solid circle, and the bright white ejecta blanket is outlined in solid black. Likewise, panel (i) shows the SSI image of Manann\'{a}n crater and the corresponding sulfuric acid, salt, and water-ice maps are shown in panels (j), (k), and (l), respectively. The extent of the dark ejecta is indicated by the solid circle, and the bright white ejecta blanket is outlined in black. The maps report the percent contribution of sulfuric acid, salts, or water-ice to the best-fit linear spectral model for each NIMS pixel. Note the different scale bars between the Pwyll and Manann\'{a}n maps. Both Pwyll and Manann\'{a}n craters and their ejecta appear to be depleted in sulfuric acid relative to nearby older terrains, suggesting that the radiolytic sulfur cycle has not yet had enough time to reach an equilibrium concentration of \ch{H2SO4} at either crater. At both Pwyll and Manann\'{a}n, the crater and dark proximal ejecta appear to be enriched in salts relative to the bright white ejecta blanket, suggesting that the dark material exhumed by the impacts is relatively salt-rich. The dark albedo and reddish coloration seen in the SSI images of both craters suggests that this salty material has already been irradiation reddened. As seen in panels (j) and (k), the nearby triple band, Belus linea, is significantly enhanced in salts and depleted in sulfuric acid relative to the background terrain. }\label{fig:maps}
\end{figure*}

Figure \ref{fig:maps} maps the pixel specific results of our linear spectral modeling analysis for hydrated sulfuric acid, the non-ice, non-sulfuric acid ``salty" component, and water-ice, respectively. The compositional trends seen in the average spectra of Figures \ref{fig:pwyll_specs} and \ref{fig:manannan_specs} are generally consistent with those seen in the compositional maps of Figure \ref{fig:maps}, although the additional spatial information in the compositional maps provides further insight. As can be seen in Figure \ref{fig:maps} (b), (f), and (j), Pwyll and Manann\'{a}n craters, their dark proximal ejecta, and their bright ray ejecta are all depleted in sulfuric acid relative to the nearby background terrains, as expected based on the average spectrum shown in Figures \ref{fig:pwyll_specs} and \ref{fig:manannan_specs}. At Pwyll, the crater and proximal ejecta have best-fit modeled sulfuric acid abundances ranging from $\sim$15-35\%, while the bright ray ejecta ranges from $\sim$20-40\%, and the older background terrain ranges from $\sim$40-55\%. There appears to be a slight gradient in the sulfuric acid abundance of the bright ray ejecta, with higher sulfuric acid concentrations farther from crater. The bright ray ejecta also appears to be relatively optically thin, with some portions of the darker background terrain showing through in Figure \ref{fig:maps} (a). This is consistent with the continuous ejecta blanket becoming progressively thinner at greater distance from the crater, and an increasing contribution to the spectrum from the underlying terrain increasing the measured sulfuric acid abundance. There are also small regions of somewhat enhanced sulfuric acid concentration (up to $\sim$40\%) within the bright ray ejecta which roughly correlate with darker patches in the SSI image where the underlying, older terrain shows through. It is also possible that sulfuric acid already present on the older terrain has been partially mixed into the bright ejecta blanket, perhaps through impact gardening. In either case, we expect that the slight increase in sulfuric acid concentration within Pwyll's bright ray ejecta, particularly in regions farther from the crater, is likely due to the older terrain showing through.  

As shown in Figure \ref{fig:maps} (j), the Manann\'{a}n crater/proximal ejecta and bright ray ejecta are also clearly depleted in sulfuric acid relative to the background terrain. The background terrain has a modeled sulfuric acid abundance which ranges from $\sim$45-65\%, while the crater/proximal ejecta show modeled sulfuric acid abundances from $\sim$35-45\% and the bright ray ejecta ranges from $\sim$35-55\%. Like at Pwyll, Manann\'{a}n's bright ray ejecta shows a stronger signature of sulfuric acid farther from the crater, where it is more optically thin, indicating there is likely to be a contribution from the underlying terrain. Additionally, the Manann\'{a}n crater and ejecta appear to be somewhat less depleted in sulfuric acid than Pwyll crater and its ejecta, relative to their respective backgrounds. This is consistent with Manann\'{a}n's older age, and thus more time for the radiolytic sulfur cycle to have produced hydrated sulfuric acid at Manann\'{a}n. It is interesting to note that the triple band, Belus linea, is also depleted in sulfuric acid (panel j). While Belus linea must be older than the Manann\'{a}n impact, it appears to be even more depleted in sulfuric acid, particularly near (235W, 7N) where there is a circular patch of low-albedo material with sulfuric acid concentrations as low as 14$^{+5}_{-5}$\%. The large amount of endogenic ``salty" material within the triple band, as seen in Figure \ref{fig:maps} (k), may inhibit the radiolytic production of sulfuric acid, although we cannot rule out the possibility of postimpact geologic processes having altered the composition of this area.
    
As can be seen in Figure \ref{fig:maps} (c) and (g), the Pwyll bright ray ejecta has a very low concentration of salts, ranging from $\sim$0-10\%, while the crater itself and dark proximal ejecta are relatively enhanced in salts, with modeled abundances between $\sim$10-25\%. Parts of the older background terrain near Pwyll also show a high concentration of salts, with abundances ranging from $\sim$10-25\%. Figure \ref{fig:maps} (k) reveals that the Manann\'{an} crater and ejecta as well as the nearby background terrain generally have a higher salt concentration than Pwyll, consistent with the average spectra shown in Figures \ref{fig:pwyll_specs} and \ref{fig:manannan_specs}. The Manann\'{a}n crater and proximal ejecta show concentrations ranging from $\sim$25-35\%. While the salt concentration of the bright ray ejecta ranges from $\sim$20-35\%, the brighter and thicker regions of the ray ejecta closest to the crater show the lowest salt abundances ($\sim$20-25\%), suggesting that the background terrain may be contributing to the measured salt abundance in the thinner or more heavily degraded regions of the bright ray ejecta. If we consider the salt abundance of the bright ray ejecta closest to the crater, the Manann\'{an} dark crater/proximal ejecta appears to be enhanced in salts relative to the bright ray ejecta. At both Pwyll and Manann\'{a}n, the low-albedo material in the crater and proximal ejecta have a higher concentration of salts than the higher-albedo material in the bright ray ejecta, suggesting that the low-albedo material thought to be excavated from Europa's subsurface likely contains endogenic salts. This is consistent with the analysis of \citet{fanale2000_TyrePwyllGalileo} who found that the visually dark and red material in the Pwyll crater floor and proximal ejecta correspond with a spectrally red unit characterized by asymmetric absorption bands. The background terrain near Manann\'{a}n is quite salty, with abundance values ranging from $\sim$30-50\%, likely due to endogenic material present within the Dyfed Regio chaos terrains. The highest salt concentrations are seen in the Belus linea triple band, with concentrations ranging from $\sim$45-55\%. We note that while there appears to be some instrumental artifacts and correlated noise in the NIMS observations (e.g. possible vertical striping in panel (g)), the magnitude of these artifacts are smaller than the contrast seen in the model abundances between the different geologic units and should therefore have a minimal effect on our scientific results.

\section{Discussion}

\subsection{Sulfuric Acid Production}\label{sec:SAdiscussion}

At both Pwyll and Manann\'{a}n, the dark material in the crater floor and proximal ejecta, and the bright ray ejecta blankets are depleted in sulfuric acid relative to nearby older terrain. Pwyll, the younger of the two craters, appears to be more significantly depleted in sulfuric acid. Assuming the background terrain concentrations represent the equilibrium concentration of sulfuric acid for the irradiation environment near each crater, this suggests that the radiolytic sulfur cycle has not yet had enough time to build up an equilibrium concentration of \ch{H2SO4} at either Pwyll or Manann\'{a}n. This is consistent with the results of \citet{hendrix2011_EuropaDiskresolvedUltraviolet}, who found that a 280 nm absorption feature attributed to \ch{SO2}, a possible byproduct of the radiolytic sulfur cycle, had a lower band strength near Pwyll with increasing band depth farther from the crater. Together, these results place a strong lower limit of the age of Pwyll and Manann\'{a}n craters on the equilibrium timescale of the sulfuric acid cycle operating on Europa's trailing hemisphere. 

While the exact ages of Pwyll and Manann\'{a}n craters remain unknown, we can use their estimated ages and make some simplifying assumptions to calculate a rough, order of magnitude estimate for the equilibrium timescale of Europa's radiolytic sulfur cycle. \citet{bierhaus2001_PwyllSecondariesOther} estimate Pwyll is $\sim$3-18 Myr old based on solar system cratering rates. However, estimates of ion sputtering erosion from \citet{paranicas2001_ElectronBombardmentEuropa} indicate that the top 1 cm of ice should be redistributed in approximately 0.6 Myr, which we would expect to degrade Pwyll's hemisphere-crossing bright rays and suggests that Pwyll may be even younger, perhaps less than 1 Myr \citep{bierhaus2009crater}. This young age is consistent, within a factor of 2, with the \citet{zahnle2003cratering} estimate that a 20 km crater should form once every $\sim$2 Myr and that Pwyll appears to be the freshest and youngest large crater on Europa. While a concrete age estimate for Manann\'{a}n crater does not exist in the literature, Manann\'{a}n is expected to be relatively young, postdates all nearby terrains including Belus linea, and likely formed under similar ice shell properties as Pwyll \citep{schenk2009impact}, suggesting it may be the second youngest crater $>$20 km on Europa. We therefore use the cratering rates of \citet{zahnle2003cratering} to roughly estimate its age to be $\sim$2 Myr older than Pwyll. This is generally consistent with its observed limited rays, which may be the remnants of a once larger ray system which have subsequently been degraded by sputtering erosion, but has not yet fully degraded. 

Considering the low end of the age estimates for Pwyll and Manann\'{a}n, we assume that Pwyll is $\sim$1 Myr old and Manann\'{a}n is $\sim$3 Myr old. We now carry out a very simple calculation to estimate a lower bound on the equilibrium timescale of the sulfuric acid cycle on Europa. As seen in Figure \ref{fig:maps}, Pwyll appears to have reached a sulfuric acid concentration which has a model abundance of $\sim$50\% the equilibrium abundance of the background terrain. If we assume that a 50\% deficit in the best-fit spectral model corresponds to a 50\% deficit in the physical concentration of sulfuric acid, then, assuming a linear growth in concentration, we would expect it to take at least 2 Myr to reach the equilibrium concentration at Pwyll. Similarly, the Manann\'{a}n crater and ejecta has a sulfuric acid abundance of $\sim$70\% of its equilibrium abundance. If it has taken $\sim$3 Myr to produce this amount of sulfuric acid, then we might expect to reach an equilibrium concentration at Manann\'{a}n after $\sim$4.3 Myr. Taking these together, we suggest that the equilibrium timescale for the radiolytic sulfur cycle on Europa's trailing hemisphere is likely to be at least a few million years, and perhaps even longer depending on the true ages of Pwyll and Manann\'{a}n craters.

It is important to note that the equilibrium timescale for the radiolytic sulfur cycle may not be consistent across Europa's trailing hemisphere. Differences in the flux of bombarding particles and changes in the surface composition may lead to differences in the equilibrium concentration of hydrated sulfuric acid and the timescale for reaching that concentration. Pwyll and Manann\'{a}n are located at similar angular distances from the trailing hemisphere apex (25\degrees and 30\degrees, respectively). We therefore expect both craters to receive a similar flux of cold sulfur ions, based on the irradiation models of \citet{cassidy2013_MagnetosphericIonSputtering}, who suggest that the flux of cold sulfur plasma to Europa's trailing hemisphere drops off by a $\cos{\theta}$ relationship, where $\theta$ is the angular distance from trailing apex. However, in our simple spectral models the background terrain near Manann\'{a}n has a higher equilibrium concentration of hydrated sulfuric acid ($\sim$50-65\%) than the terrain near Pwyll ($\sim$40-55\%), consistent with the overall distribution of sulfuric acid hydrate found by \citet{ligier2016_VLTSINFONIOBSERVATIONS}. This suggests that the equilibrium concentration of sulfuric acid, and likely the timescale to reach that concentration, may depend on more than the sulfur ion flux. For example, the flux of 0.1–25 MeV electrons, as estimated by \citet{paranicas2009_EuropaRadiationEnvironment}, is expected to be higher at Manann\'{a}n than Pwyll and may help explain the higher equilibrium concentration of sulfuric acid near Manann\'{a}n. This is consistent with the hypothesis proposed by \citet{dalton2013_ExogenicControlsSulfuric} that the production and steady-state concentration of hydrated sulfuric acid may depend on both the concentration of iogenic sulfur ions and the flux of energetic electrons.

Laboratory estimates for the sulfur cycle equilibrium timescale on Europa's trailing hemisphere are typically $\sim$10$^4$ yr \citep[e.g.][]{carlson2002_SulfuricAcidProduction, strazzulla2007_HydrateSulfuricAcida, ding2013_ImplantationMultiplyCharged}. However, our results suggest that it may take several orders of magnitude longer to reach an equilibrium concentration of \ch{H2SO4} on Europa. It is very unlikely that both Pwyll and Manann\'{a}n craters are young enough to be consistent with a sulfuric acid equilibrium timescale of a few thousand years, so we must conclude that the equilibrium timescale for the radiolytic sulfur cycle is significantly underestimated by these laboratory estimates. \citet{strazzulla2007_HydrateSulfuricAcida} and \citet{ding2013_ImplantationMultiplyCharged} calculate these timescales using their experiment-derived reaction rates, estimates for the column density of hydrated \ch{H2SO4} molecules in the observable surface layers on Europa from \citet{carlson1999_SulfuricAcidEuropa}, and estimates for the KeV-MeV sulfur ion flux from \citet{cooper2001_EnergeticIonElectron} and \citet{dalton2013_ExogenicControlsSulfuric}, respectively. These calculations would therefore underestimate the equilibrium timescale if the actual sulfuric acid column density at Europa was higher than that estimated by \citet{carlson1999_SulfuricAcidEuropa}, or if the real sulfur ion flux is lower than that estimated by \citet{cooper2001_EnergeticIonElectron} and \citet{dalton2013_ExogenicControlsSulfuric}. However, it is difficult to achieve a 2 order of magnitude increase in the equilibrium timescale by simply adjusting the sulfuric acid column density and sulfur ion flux, which suggests that understanding the discrepancy between our observational constraints and the equilibrium timescales derived from laboratory experiments may require more careful analysis.

Laboratory experiments must necessarily replicate Europa's irradiation environment over much shorter timescales, perhaps reaching an equivalent irradiation dose of a few thousand to a million years on Europa in a matter of hours. Additionally, most laboratory experiments are performed for a single bombarding particle and energy, while various species and energies are acting simultaneously at Europa. Indeed, laboratory experiments have demonstrated that the specific intermediary products and steady-state concentration of \ch{H2SO4} may depend strongly on the temperature, energies, and projectiles involved \citep[e.g.][]{moore2007_RadiolysisSO2H2S, strazzulla2007_HydrateSulfuricAcida, loeffler2011_RadiolysisSulfuricAcid}. Thus, using a laboratory-derived reaction rate for a single bombarding particle and energy may not provide a robust estimate for the actual equilibrium timescale at Europa, which may help explain the discrepancy between our observational constraints and the estimates of \citet{carlson2002_SulfuricAcidProduction}, \citet{strazzulla2007_HydrateSulfuricAcida}, and \citet{ding2013_ImplantationMultiplyCharged}. Additionally, while radiolysis is responsible for the formation of \ch{H2SO4} on Europa, it can also destroy these molecules once formed. The balance of the production and destruction of hydrated \ch{H2SO4} on Europa is not well known, but is likely to be heavily particle and energy dependent. If \ch{H2SO4} is more readily destroyed within the complex irradiation environment on Europa's trailing hemisphere than in laboratory experiments with single-particle bombardment, then these destructive processes may act to prolong the timescale over which sulfuric acid reaches a steady-state concentration on Europa. Additionally, it is possible that thermal processes which are unimportant at laboratory timescales may become important on Europa, and could serve to slow down the equilibrium timescale of the radiolytic sulfur cycle. Indeed, \citet{tribbett2022_ThermalReactionsH2S} demonstrated that \ch{H2O + H2S + O3} ice mixtures at Europa temperatures can undergo thermal oxidation reactions, even on laboratory timescales, which suggests that thermal reactions may affect both the steady-state composition and intermediary products of the radiolytic sulfur cycle.

\subsection{Irradiation-induced Reddening of Salts}

We find that the dark, red material in the crater and proximal ejectas of both Pwyll and Manann\'{a}n contain higher salt abundances than the surrounding bright ray ejecta. This dark material was likely excavated from Europa's subsurface during impact \citep{moore1998_LargeImpactFeatures, schenk2009impact}, and is therefore expected to contain endogenic material which may have been further concentrated via preferential vaporization of ice during the impact process \citep{tomlinson2022_CompositionPossibleOrigins}. Indeed, \citet{fanale2000_TyrePwyllGalileo} analyzed the NIMS observations shown in Figure \ref{fig:maps} (f), (g), and (h) and found that the visually dark and red material in the crater floor and proximal ejecta have a spectral signature dominated by asymmetric absorption bands, which they broadly attributed to non-ice hydrated material. In the framework of our spectral modeling analysis, these asymmetric bands can be explained by a variety of hydrated salts but are not well reproduced by a simple combination of water-ice and hydrated sulfuric acid. Endogenic, presumably salty material within Europa's other geologically young features like the chaos terrains, bands, and ridges become progressively darker and redder toward the trailing hemisphere apex, potentially due to irradiation-induced reddening of salts. We expect that the endogenic salts within the Pwyll and Manann\'{a}n craters and proximal ejecta have also been darkened and reddened via irradiation, based on their low-albedo and visually red coloration in the Galileo/SSI false-color images. If this is the case, the timescale for irradiation-induced reddening of salts on Europa's trailing hemisphere is apparently shorter than the ages of Pwyll and Manann\'{a}n craters. Taking the lower end of the age estimates for Pwyll, as in Section \ref{sec:SAdiscussion}, we can place an approximate upper limit of $\sim$1 Myr on the irradiation-induced reddening timescale of salts on Europa's trailing hemisphere.

\section{Conclusions}

We have performed a simple linear spectral modeling analysis on the available high-spatial-resolution Galileo/NIMS observations of Pwyll and Manann\'{a}n craters in order to constrain the irradiation timescales of the radiolytic sulfur cycle and irradiation-induced reddening of salts operating on Europa's trailing hemisphere. Both Pwyll and Manann\'{a}n craters and their ejecta are depleted in hydrated sulfuric acid relative to nearby older terrain. The radiolytic sulfur cycle has apparently not yet had enough time to build up an equilibrium concentration of \ch{H2SO4} at either crater, placing a strong lower limit of the age of these craters on its equilibrium timescale. Based on these NIMS observations and a rough estimate for the ages of Pwyll and Manann\'{a}n, we estimate that the equilibrium timescale for the production of hydrated sulfuric acid on Europa's trailing hemisphere is likely to be at least a few million years. This is several orders of magnitude longer than previous estimates of $\sim$$10^4$ yr based on laboratory experiments and suggests that simultaneous production and destruction of \ch{H2SO4} and perhaps thermal processes may be important for understanding the radiolytic sulfur cycle on Europa.

We also find that the dark and red material in the craters and proximal ejecta of Pwyll and Manann\'{a}n contain an abundance of hydrated salts, presumably endogenic in origin. This salty material has apparently already been darkened and reddened via irradiation. We are therefore able to place an upper limit of the age of Pwyll crater, perhaps $\sim$1 Myr, on the timescale for irradiation-induced reddening of salts on Europa's trailing hemisphere.

The authors would like to thank Samantha K. Trumbo, William T. Denman, and Matthew Belyakov for helpful discussions.

%
\vspace{5mm}

\software{Cartopy \citep{Cartopy},
          GeoPandas \citep{geopandas},
          SciPy \citep{2020SciPy-NMeth},
          Shapely \citep{shapely2023}
          }



\bibliography{bibliography}{}
\bibliographystyle{aasjournal}

\end{document}